# ASYMMETRIC OPTICAL AMPLIFIER BASED ON PARITY-TIME SYMMETRY

Rujiang LI, Pengfei LI, Lu LI

Institute of Theoretical Physics, Shanxi University, Taiyuan 030006, China
E-mail: llz@sxu.edu.cn

We design and optimize a waveguide structure consisted of three segments of a parity-time symmetric coupler. It can amplify the difference mode jointly with the attenuation of the sum mode, so it performs an asymmetric optical amplifier. Also, the dependence of the amplification factor for the difference mode and the attenuation factor for the sum mode on the gain/loss parameter is illustrated numerically.

## 1. INTRODUCTION

Since the concept of parity-time (PT) symmetry was proposed in the framework of quantum mechanics [1], we have seen a rapid development because it can exhibit entirely real eigenvalue spectra even for the non-Hermitian Hamiltonian with complex potential $V(\mathbf{r}) = V^*(-\mathbf{r})$ [1–3]. Although the impact of PT-symmetry in quantum mechanics is still debated, it has been shown that, in optics, PT-related notions can be implemented in PT-symmetric couplers and double channel waveguides [4–9] and PT-symmetric optical lattices [10–19], and the experimental observations have been demonstrated [20, 21]. Thus, optics can provide a fertile ground to investigate PT-related beam dynamics including the non-reciprocal responses, the power oscillations, and the optical transparency. Recently, based on the properties in the PT-symmetry breaking region, different optical components are springing up both theoretically and experimentally [22–24]. Since the reciprocity theorem of linear systems, the asymmetric optical components are mainly performed in the nonlinear regime which relies strictly on the power of the incident beam [25]. In this paper, we propose a simple waveguide structure consisted of three segments of linear PT-symmetric coupler which performs below the PT-symmetry breaking point. This component can amplify the difference mode jointly with the attenuation of the sum mode, so it performs an *asymmetric optical amplifier*.

## 2. MODEL AND REDUCTIONS

In the context of the paraxial theory of diffraction by involving symmetric index guiding and an antisymmetric gain/loss profile, the electric field envelope obeys a normalized complex Schrödinger equation as follows

$$i\frac{\partial \varphi}{\partial \zeta} + \frac{\partial^2 \varphi}{\partial \xi^2} + V(\xi)\varphi = 0 . \tag{1}$$

Here $\zeta = z/2kx_0^2$ is a scaled propagation distance and $\xi = x/x_0$ is a dimensionless transverse coordinate, where $x_0$ is an arbitrary spatial scale and $k = 2\pi n_0/\lambda_0$ is the wave number with $n_0$ being the background refractive index and $\lambda_0$ being the wavelength of the optical source generating the beam. The function $V(\xi) = 2k^2 x_0^2 [n(\xi) - n_0]/n_0 \equiv U(\xi) + iW(\xi)$ represents the normalized complex index distribution that satisfies the PT condition. Here we consider the PT-symmetric double-channel waveguide structure, in which the refractive index distribution is of the form $U(\xi) = U_1(\xi) + U_2(\xi)$ with $U_1(\xi)$ and $U_2(\xi)$ taking $U_0$ as $-L_0/2 - D_0 < \xi < -L_0/2$ and $L_0/2 < \xi < L_0/2 + D_0$, respectively, and the gain/loss distribution is of the form $W(\xi) = W_1(\xi) + W_2(\xi)$ with $W_1(\xi)$ and $W_2(\xi)$ taking $W_0$ within $-L_0/2 - D_0 < \xi < -L_0/2$ and $L_0/2 < \xi < L_0/2 + D_0$, respectively, and otherwise $U_1(\xi) = U_2(\xi) = W_1(\xi) = W_2(\xi) = 0$. Here $U_0$ is the modulation depth of the refractive index, $W_0$ is the dimensionless gain/loss parameter, and $L_0$ and $D_0$ correspond to the scaled separation and width of channels, respectively. Eq. (1) can describe beam propagation in PT-symmetric double-channel waveguide. The optical modes and the beam dynamics in such coupler have been studied widely [4–9].

In order to investigate the dynamics of beam propagation in such waveguide, we will employ the *coupled-mode approach*. In this case, the solution of Schrödinger equation can be expressed as the superposition of local modes of individual channels without gain and loss, i.e.

$$\varphi(\zeta, \xi) = \left[ a(\zeta) u_1(\xi) + b(\zeta) u_2(\xi) \right] \exp(i\beta\zeta). \tag{2}$$

Here $u_j(\xi)$ satisfies $d^2 u_j / d\xi^2 + U_j(\xi) u_j = \beta u_j$, $j = 1, 2$. Substituting Eq. (2) into Eq. (1), multiplying by $u_2(-\xi)$ and $u_1(-\xi)$ respectively, and integrating over $\xi$ yield the following set of equations

$$i \frac{da}{d\zeta} + (\delta - i\gamma) a + (\kappa - i\sigma) b = 0, \tag{3}$$

$$i \frac{db}{d\zeta} + (\delta + i\gamma) b + (\kappa + i\sigma) a = 0. \tag{4}$$

Here $\delta = (I_{12} J_{121} - I_{11} J_{122})/\Delta$, $\gamma = I_{12}(K_{121} - K_{121})/\Delta$, $\kappa = (I_{12} J_{122} - I_{11} J_{121})/\Delta$ and $\sigma = I_{11}(K_{112} - K_{121})/\Delta$ with $\Delta = I_{12}^2 - I_{11}^2$, $I_{mj} = \int_{-\infty}^{+\infty} u_m(\xi) u_j(-\xi) d\xi$, $J_{mjk} = \int_{-\infty}^{+\infty} U_m(\xi) u_j(\xi) u_k(-\xi) d\xi$ and $K_{mjk} = \int_{-\infty}^{+\infty} W_m(\xi) u_j(\xi) u_k(-\xi) d\xi$, and we have used the relations $I_{22} = I_{11}$, $I_{21} = I_{12}$, $J_{212} = J_{121}$, $J_{122} = J_{211}$, $K_{221} = K_{112}$, $K_{212} = K_{121}$. In Eqs. (3) and (4) $\delta$ represents the phase shift induced by the neighbor refractive index distribution, $\gamma$ is a scaled effective gain/loss coefficient, while $\kappa$ and $\sigma$ are the real and imaginary part of the scaled coupling coefficient, respectively.

### 3. OPTICAL AMPLIFIER

Here we provide a new structure consisted of three segments of double-channel waveguides with equal amount of gain and loss changing alternatively, forming a double-channel structure of gain-loss-gain (GLG) channel and loss-gain-loss (LGL) channel, as shown in Fig. 1. In the following, we discuss beam dynamics in each segment waveguide by solving Eqs. (3) and (4).

It can be shown that there exists a critical point $\gamma = \sqrt{\kappa^2 + \sigma^2}$ such that below this critical point, i.e., $\gamma < \sqrt{\kappa^2 + \sigma^2}$ the eigenmodes are given by $\left( g e^{\mp i\alpha}, \pm e^{\pm i\alpha} \right)^T$ with the corresponding eigenvalues being $\lambda_\pm = \delta \pm \omega$, where $\omega = \sqrt{\kappa^2 + \sigma^2 - \gamma^2}$, $\sin 2\alpha = \gamma / \sqrt{\kappa^2 + \sigma^2}$, $g = \sqrt{(\kappa - i\sigma)/(\kappa + i\sigma)}$ with $|g|^2 = 1$, and the superscript 'T' represents the transpose. Thus, the solutions of Eqs. (3) and (4) in each segment can be written as

$$\begin{pmatrix} a(\zeta) \\ b(\zeta) \end{pmatrix} = U(\zeta) \begin{pmatrix} a_0 \\ b_0 \end{pmatrix}, \tag{5}$$

where the matrix $U(\zeta)$ is given by

$$U(\zeta) = \frac{e^{i\delta\zeta}}{\cos(2\alpha)} \begin{pmatrix} \cos(\omega\zeta - 2\alpha) & ig\sin(\omega\zeta) \\ ig^*\sin(\omega\zeta) & \cos(\omega\zeta + 2\alpha) \end{pmatrix}, \tag{6}$$

and $(a_0, b_0)^T$ represents the initial state in each segment waveguide. It should be pointed out that in the second segment waveguide $\alpha$ and $g$ in the matrix (6) are replaced by $-\alpha$ and $g^*$, respectively, because of the changes of the sign of the parameters $\gamma$ and $\sigma$. From Eq. (5), it can be seen that the power $P(\zeta) = |a(\zeta)|^2 + |b(\zeta)|^2$ oscillates with the period $\pi/\omega$, which corresponds to the coupling length $\pi/\kappa$ in the absence of the gain/loss effect [5, 21].

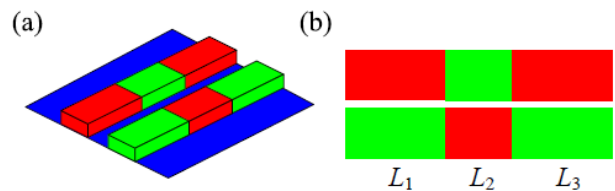

Fig. 1 – a) Schematic illustration of the waveguide structure. Waveguide are color-coded according to the gain or loss: red for gain and green for loss; b) the platform of the structure, where $L_1$, $L_2$ and $L_3$ denote the length of each segment waveguide, respectively.

Our aim is to optimize the length of the three segments of waveguides such that it can exhibit a feature of the asymmetric amplification. First, we consider the difference of the eigenmodes as an input state

$$(a(0), b(0))^T = (-ig\sin\alpha, \cos\alpha)^T, \tag{7}$$

which corresponds to the state $(0,1)^T$ in the absence of the gain/loss effect (i.e., $\gamma = 0$ and $\sigma = 0$). Note that in our design process we take the output power in each segment waveguide as large as possible.

Thus, taking Eq. 7 as the initial state in Eq. (5) yields the power $P_1(\zeta) = 1 - \sin(2\alpha)\sin(2\omega\zeta)$ in the first segment waveguide, from which one can find that it reaches the maximum value at $\zeta = 3\pi/(4\omega)$. Thus we take $L_1 = 3\pi/(4\omega)$, and the corresponding output state at $\xi = L_1$ is

$$\begin{pmatrix} a(L_1) \\ b(L_1) \end{pmatrix} = \sqrt{P_1(L_1)} \frac{e^{i\delta L_1}}{\sqrt{2}} \begin{pmatrix} ig \\ -1 \end{pmatrix}. \tag{8}$$

In the second segment waveguide, taking Eq. (8) as the initial state in Eq. (5) yields the power $P_2(\zeta) = P_1(L_1)[1 + A_2\sin^2(\omega\zeta)]$ with $A_2 = [2\sin^2(2\alpha) + (g^2 + g^{*2})\sin(2\alpha)]/\cos^2(2\alpha)$. One can find that $P_2(\zeta)$ reaches the maximum value at $\zeta = \pi/(2\omega)$, so we take $L_2 = \pi/(2\omega)$. In this case, the corresponding output state at $\zeta = L_1 + L_2$ is of the form

$$\begin{pmatrix} a(L_1 + L_2) \\ b(L_1 + L_2) \end{pmatrix} = F_2 \begin{pmatrix} -ig^* - ig\sin(2\alpha) \\ -g^2 - \sin(2\alpha) \end{pmatrix}, \tag{9}$$

where $F_2 = \sqrt{P_2(L_2)/N_2} \exp[i\delta(L_1 + L_2)]$ with $N_2 = 2[g^2 + \sin(2\alpha)][g^{*2} + \sin(2\alpha)]$ being a normalized factor.

Finally, taking Eq. (9) as the initial state in Eq. (5), it can be shown that the power in the third segment waveguide is of the form $P_3(\zeta) = P_2(L_2)[1 + 2A_3\sin^2(\omega\zeta)]$, where $A_3 = \{N_2 \sin^2(2\alpha) + \sin(2\alpha)[g^{*2} + \sin(2\alpha)]^2 + \sin(2\alpha)[g^2 + \sin(2\alpha)]^2\}/[N_2 \cos^2(2\alpha)]$, and we obtain the final output state

$$a(L) = F_3 \{-ig[g^2 + \sin(2\alpha)]\sin(\omega L_3) - ig[g^{*2} + \sin(2\alpha)]\cos(\omega L_3 - 2\alpha)\},$$
$$b(L) = F_3 \{[g^{*2} + \sin(2\alpha)]\sin(\omega L_3) - [g^2 + \sin(2\alpha)]\cos(\omega L_3 + 2\alpha)\}, \quad (10)$$

where $L = L_1 + L_2 + L_3$ and $F_3 = \sqrt{P_3(L_3)/N_3}\exp(i\delta L)$, with

$$N_3 = N_2 \cos^2(2\alpha) + 2\{N_2 \sin^2(2\alpha) + \sin(2\alpha)[g^2 + \sin(2\alpha)]^2 + \sin(2\alpha)[g^{*2} + \sin(2\alpha)]^2\}\sin^2(\omega L_3) \quad (7)$$

being a normalizing factor. It should be pointed out that, as an amplification component, the output state at $\xi = L$ should be of the form of the difference mode 7. Thus one have to determine the length $L_3$ of the third segment waveguide such that it satisfies the equation $a(L)/b(L) = -ig\tan\alpha$. However, in general, this equation has no any real solution. Fortunately, in the special case when $g = 1$, this equation can be reduced to $\tan(\omega L_3) = -1$, from which one can find $L_3 = 3\pi/(4\omega)$. Thus the power of the output state in the form of the difference mode can be written as $P_{am} = [1 + \sin(2\alpha)]^4/\cos^4(2\alpha)$, which means that the difference mode is amplified.

For the same conditions, we also take the sum of the eigenmodes as an input state

$$(a(0), b(0))^T = (g\cos\alpha, i\sin\alpha)^T, \quad (11)$$

which corresponds to the state $(1,0)^T$ in the absence of the gain/loss effect. The power of the output state in the form of the sum mode is $P_{at} = [1 - \sin(2\alpha)]^4/\cos^4(2\alpha)$. This means that the sum mode is attenuated.

The above results reveal that the waveguide structure presented here plays the role of an *asymmetric amplifier*, which is similar to the differential amplifier in electronic circuits [26]. Here, $P_{am}$ and $P_{at}$ represent the amplification and the attenuation factor, respectively, since the incident power is normalized to 1.

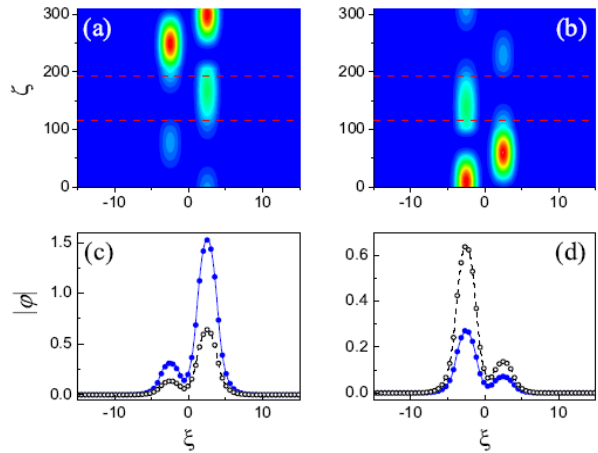

Fig. 2 – The evolution plots of the beam for $W_0 = 0.01$, (a) and (b); the amplitude distributions of the input (black dashed) and the output (blue solid) states, (c) and (d). Here (a) and (c) denote the results for the difference mode, while (b) and (d) denote the results for the sum mode. The lines and the circles are the theoretical results for $g = 1$ and the numerical results from Eq. (1), respectively. The red dashed lines indicate the sideline of the three segments of waveguide.

It should be noted that the above results hold only for $g = 1$, which can not be exactly satisfied in real applications. However, a lot of simulations shown that $\sigma$ is a small quantity compared with $\kappa$, thus $g \approx 1$. As an example, here we performed the evolutions of the difference mode and the sum mode by simulating numerically Eq. (1), where $n_0 = 3.2496$ (the cladding) and $n_1 = 3.2683$ (the core) [20], which correspond to $U_0 = 2$ for the wavelength $\lambda_0 = 1.55\,\mu m$ and $x_0 = 1\,\mu m$, and $L_0 = 2$ and $D_0 = 3$, respectively. In this case $g = 1 - 0.0167i$, and the total length of the amplifier is $2\pi/\omega = 8.14$ mm. Figure 2 presents the evolution plots of the difference and the sum mode, and the corresponding amplitude distributions of the input and the output states for $W_0 = 0.01$, respectively. From Fig. 2, one can see that the difference mode experiences oscillatory amplification between GLG and LGL channels, and eventually evolves into an amplified output state, in which the powers of the input and the output states mainly focus on the LGL channel. While for the sum mode, it is attenuated and the powers of the input and the output states mainly focus on the GLG channel. Also, it is shown that the theoretical result agrees quite well with the numerical simulation. In order to understand the influence of the asymmetric amplification on the gain/loss parameter, the dependence of the amplification factor $P_{am}$ and the attenuation factor $P_{at}$ on the gain/loss parameter $W_0$ is plotted in Fig. 3, which means that the asymmetric amplification effect is more evidence for larger $W_0$ below the symmetry-breaking point ($W_0 \approx 0.0024$). It is worth to note that this asymmetric amplification only exists for the PT-symmetric case, whereas for the conventional optical systems with Hermitian Hamiltonians, the light works in a reciprocal behavior [6].

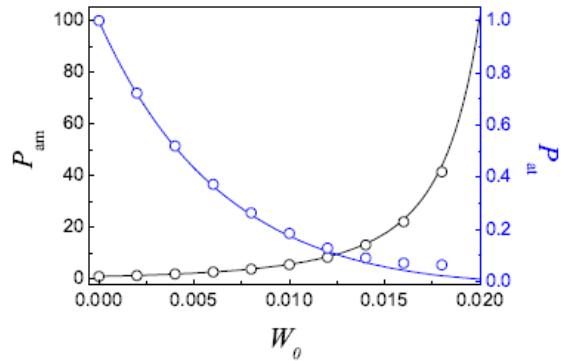

Fig. 3 – The dependence of the amplification factor $P_{am}$ (black) and the attenuation factor $P_{at}$ (blue) on the gain/loss parameter $W_0$. The solid lines denote the theoretical results for $g = 1$, while the open circles denote the numerical results.

## 4. CONCLUSIONS

In conclusion, we have addressed a new way to construct an asymmetric amplifier based on PT-symmetric waveguides. By employing the coupled-mode theory, the structure of the asymmetric amplifier has been optimized. The results have shown that the difference mode can be amplified and at the same time, the sum mode is attenuated. These results may provide important hints to the design and optimization of other practical optical components which can be used in all-optical signal processing devices and systems.

## ACKNOWLEDGEMENTS

This research is supported by the National Natural Science Foundation of China grant 61078079 and the Shanxi Scholarship Council of China Grant No. 2011-010.